%% file: main.tex
\documentclass[lettersize,journal]{IEEEtran}
    \usepackage{siunitx}
    \usepackage{xpatch}
    \usepackage{hyperref}       
    \usepackage{float}
    \usepackage{xcolor}
    \usepackage{ragged2e}
    \usepackage{fixltx2e}
    \usepackage{siunitx}
    \usepackage{tabularx}
    \usepackage{listings}
    \usepackage{arydshln}
    \usepackage{indentfirst}
    
    \usepackage[backend=biber, style=numeric, sorting=none]{biblatex}
    \definecolor{codegreen}{rgb}{0,0.6,0}
    \definecolor{codegray}{rgb}{0.5,0.5,0.5}
    \definecolor{codepurple}{rgb}{0.58,0,0.82}
    \definecolor{backcolour}{rgb}{0.95,0.95,0.92}

    \lstdefinestyle{mystyle}{
        backgroundcolor=\color{backcolour}, 
        commentstyle=\color{codegreen},
        keywordstyle=\color{magenta},
        numberstyle=\tiny\color{codegray},
        stringstyle=\color{codepurple},
        basicstyle=\ttfamily\footnotesize,
        breakatwhitespace=false,         
        breaklines=true,                 
        captionpos=b,                    
        keepspaces=true,                 
        numbers=left,                    
        numbersep=5pt,                  
        showspaces=false,                
        showstringspaces=false,
        showtabs=false,                  
        tabsize=2
    }
    
    \IEEEoverridecommandlockouts
      \usepackage{tipa}
      \usepackage[pdftex]{graphicx}
    	\usepackage{epsfig}
    	\usepackage{epstopdf}
    
    \usepackage{hyperref}
    \hypersetup{
      pdfencoding=unicode,
      pdfstartview=FitH,
      pdfpagemode=UseNone,
      colorlinks=true,
      linkcolor=blue,
      urlcolor=blue,
      citecolor=blue
    }

\usepackage{etoolbox}
\makeatletter
\patchcmd{\@makecaption}
  {\scshape}
  {}
  {}
  {}
\makeatother

\begin{document}
\input{Sections/Title}

\author{
    P. S. Mbonimpa$^{1}$, D. Tuyizere$^{1}$ , A. A. Biyabani$^{1}$, O. K. Tonguz$^{2}$ \\
     \{pmbonimp, dtuyizer\}@alumni.cmu.edu, \{ab3x, tonguz\}@andrew.cmu.edu \\
    $^{1}$ Carnegie Mellon University - Africa, Kigali, Rwanda \\
     $^{2}$ Carnegie Mellon University, Pittsburgh, United States\\
}




\maketitle
\input{Sections/Abstract}


\input{Sections/introduction}
\input{Sections/related_work}

\input{Sections/Approach}

\input{Sections/results}

\input{Sections/Deployment_On_Edge}
\input{Sections/Conclusion}

\printbibliography

\end{document}

%% file: Sections/Title.tex
\title{Edge-Based Speech Transcription and Synthesis for Kinyarwanda and Swahili  Languages 
}

%% file: Sections/Abstract.tex
\begin{abstract}
This paper presents a novel framework for speech transcription and synthesis, leveraging edge-cloud parallelism to enhance processing speed and accessibility for Kinyarwanda and Swahili speakers. It addresses the scarcity of powerful language processing tools for these widely spoken languages in East African countries with limited technological infrastructure. The framework utilizes the Whisper and SpeechT5 pre-trained models to enable speech-to-text (STT) and text-to-speech (TTS) translation. The architecture uses a cascading mechanism that distributes the model inference workload between the edge device and the cloud, thereby reducing latency and resource usage, benefiting both ends. On the edge device, our approach achieves a memory usage compression of 9.5\% for the SpeechT5 model and 14\% for the Whisper model, with a maximum memory usage of 149 MB. Experimental results indicate that on a 1.7 GHz CPU edge device with a 1 MB/s network bandwidth, the system can process a 270-character text in less than a minute for both speech-to-text and text-to-speech transcription. Using real-world survey data from Kenya, it is shown that the cascaded edge-cloud architecture proposed could easily serve as an excellent platform for STT and TTS transcription with good accuracy and response time. 
\end{abstract}

%% file: Sections/introduction.tex
\section{\MakeUppercase{Introduction }}\label{sec:compAnalysis}

In today’s digital age, the need for accurate and efficient speech transcription and synthesis models has been increasing rapidly. These models play an important role in a variety of applications, such as learning new language(s), accessibility tools for people with difficulties in reading and hearing, as well as automated voice assistants \cite{waliaula2023}. Kinyarwanda and Swahili are two of the local languages spoken in East Africa. While Swahili is the most widely spoken language in Eastern Africa, the speakers range from 60 million to over 150 million \cite{lisanza2021swahili}. Swahili serves as the national language of four African nations: Tanzania, Kenya, Uganda, Rwanda, and the Democratic Republic of the Congo. On the other hand, Kinyarwanda is the national language of Rwanda, spoken by approximately 24 million people in Rwanda and beyond \cite{RwandaPop}. 

Rwanda and Kenya are among the developing countries in the East African Community (EAC) in the Eastern part of Africa. Although this is the case, there is a significant scarcity of speech transcription and synthesis models for such languages, mainly due to a lack of strong community-based research initiatives to tackle existing computing-related challenges \cite{Metalom_Fendji_Yenke_2022b}. In addition, there are huge, growing technology-backed services in which the availability of speech transcription and synthesis models would not only boost innovation but also contribute to research. Furthermore, many people in such countries have very little access to the Internet and they own devices with constrained computational resources. One of the challenges in this area is having models designed to run seamlessly on such devices that can perform locally at an optimal level while requesting further processing from the cloud only when the local computation does not meet important performance criteria such as accuracy and/or the time required to do text-to-speech  (TTS) or speech-to-text translation (STT).

In the past, different research efforts have been reported to develop Edge-Based speech transcription and synthesis models for languages, including under-resourced ones. For instance, EfficientSpeech \cite{Atienza} provides an On-Device text-to-speech model designed specifically for edge devices. This model aims to provide natural-sounding speech synthesis while being efficient enough to run on a local device such as a smartphone. To name a few, other approaches such as Tacotron \cite{wang2017synthesis}, Tacotron 2 \cite{Shen}, and WaveGlow \cite{WaveGlow}, have pushed the boundaries of end-to-end generative text-to-speech models. 

On the other hand, edge-based speech-to-text approaches such as 'Streaming End-to-End Speech Recognition for Mobile Devices'\cite{Anjuli} and 'Speech Recognition and Speech Synthesis Models for Micro Devices'\cite{eichner00_icslp} try to integrate the speech recognition models at the edge.  Although this is the case, there is still a huge gap in the efficiency and computation-aware intelligence of the deployed architecture models, taking into consideration the computation capacity of the edge device as well as the efficiency in relying on the cloud when support is needed. In addition, none of the edge-specific approaches tackle specifically low resource languages, such as Kinyarwanda and Kiswahili, which is an important shortcoming. 

To address this shortcoming, in this paper, we present a novel, computationally efficient cascading approach for edge-cloud speech transcription and synthesis, specifically designed for Kinyarwanda and Swahili languages. This model and the underlying mechanism was inspired by the approach proposed in \cite{Leroux2017TheCN}, a cascading neural network to efficiently couple the computation between models deployed at the edge as well as the cloud. In a nutshell, a cascading neural network is capable of distributing a deep neural network between a local device and the cloud while keeping the required communication network traffic to a minimum. The network begins processing on the constrained device and only relies on the remote part (i.e., the Cloud) when the local part does not provide an accurate or fast enough result. For instance, in Text-to-Speech (TTS), when the processed output voice after the edge processing is very noisy, it might be necessary to send the internal representation to the Cloud. Similarly, in Speech-to-Text (STT), when the input voice at the edge is noisy, its internal representations are sent to the Cloud for better processing. We present the results of encoder-decoder models for both speech-to-text as well as text-to-speech for both Kinyarwanda and Swahili. 

The primary contributions of this paper are as follows:

\begin{enumerate}

\item We introduce an efficient edge-cloud cascading mechanism that processes speech data on local devices efficiently and escalates to Cloud-based resources when necessary.

\item We present a practical deployment scheme for the speech-to-text and text-to-speech models on both the edge and the cloud devices, thus enhancing the compute capability and scalability of TTS and STT transcription.

\item Using the existing data on smartphones and Internet connectivity in Eastern Africa, we show that the proposed hybrid edge-cloud scheme can provide excellent performance for TTS and STT transcription services in terms of accuracy and response time.


\end{enumerate}

The remainder of the paper is organized as follows. Section \ref{sec:relatedwork} examines the related work. Section \ref{sec:approach}  gives an overview of the approach used, focusing on the architecture of the models considered. Section \ref{Model Training} shows the details of the model training conducted and the results obtained. Section \ref{sec:results} presents the numerical results of the experiments and a qualitative discussion of the findings while Section \ref{sec:deployment} describes deployment on the edge. Finally, Section \ref{sec:futurework} concludes the paper and discusses possible future work.

%% file: Sections/related_work.tex
\section{\MakeUppercase{Related Work }}\label{sec:relatedwork}

Recent advances in speech-to-text (STT) and text-to-speech (TTS) technologies depended on the significant progress made in machine learning and deep learning architectures. These advances emphasize the emerging challenges and opportunities in deploying these advanced models on edge devices, particularly for under-resourced languages like Kinyarwanda and Swahili. This section presents different approaches that cover STT and TTS technologies while targeting edge device computation.

\subsection{Speech-To-Text (STT) models}
In recent years, STT models have made significant strides, thanks to advances in deep neural networks. These models aim to transcribe spoken language into written text, with applications ranging from transcription services to voice assistants.  
The survey conducted in \cite{Rohit} provided a synopsis of speech-to-text models that incorporated deep neural networks. Among the various architectures present today, covering various aspects of this task including modeling, training, encoding, and decoding, transformer-based models, such as those presented in \cite{Conformer}, \cite{Guo}, \cite{Gabriel}, \cite{HuBERT} have emerged as state-of-the-art, achieving impressive word error rates (WER) on challenging datasets like Librispeech. They reduced the WER trend to 3.7\% - 1.8\% in transcribing the Librispeech dataset.   
Deployment of the models mentioned above critically requires compute capability considerations for the platform. Taking the Whisper model, introduced in \cite{Alec}) as an example, through varying the number of attention layers, five types of the model were proposed: tiny (39 million parameters), base (74 million parameters), small (244 million parameters), medium (769 million parameters) and large (1550 million parameters).  With such deep neural network models, tradeoffs between model size and accuracy need to be examined, per the specifications of the targeted edge device(s), as the results shown in \cite{Alec} showed that smaller models had poorer performance. However, network topologies that allow distributing a large model across more than a single device would enable workload sharing and maintain accuracy while meeting edge devices’ constraints.

\subsection{Text-To-Speech (TTS) models}
TTS systems have witnessed significant advancements in recent years, driven by the continuous development of deep learning architectures and training methodologies. 

One approach to achieving speech synthesis involves leveraging neural network architectures with a focus on maximizing audio quality. One of the current state-of-the-art approaches is FastSpeech 2 \cite{Ren}. FastSpeech 2 is an end-to-end text-to-speech synthesis model that focuses on generating non-autoregressive mel-spectrograms directly from text. FastSpeech 2 has been recognized for its ability to  produce high-quality speech output efficiently. Despite its remarkable capabilities in generating natural-sounding speech, its computational cost and memory footprint limit its deployment on devices with limited resources. 
Recognizing the limitations of high-resource models, researchers are actively exploring alternative strategies for building efficient and resource-constrained TTS systems.  EfficientSpeech \cite{Rowel} introduces a novel TTS model designed specifically for edge devices. This model aims to provide natural-sounding speech synthesis while being efficient enough to run on-device. EfficientSpeech joins a sequence of advancements in TTS technology, such as Tacotron \cite{Wang}, Tacotron 2 \cite{Jonathan}, and WaveGlow \cite{Ryan}, which have pushed the boundaries of text-to-speech models.

The development of EfficientSpeech aligns with the trend towards on-device processing highlighted in studies like "Streaming End-to-End Speech Recognition for Mobile Devices" \cite{Yanzhang} and "Speech  Recognition and Speech Synthesis Models for Micro Devices" \cite{Asante_Imamura_2019b}. These works emphasize the importance of deploying speech recognition and synthesis models on resource-constrained devices like mobile phones and microcontrollers, enabling applications that require low latency and privacy preservation. LightGrad \cite{Chen_2023} takes this concept a step further by adopting a non-autoregressive approach that utilizes a lightweight U-Net architecture and streaming inference to achieve even lower latency. TTS technology is not limited to well-resourced languages. LRSpeech \cite{Jin} addresses the challenge of building TTS systems for languages with limited data availability. This model leverages pre-training on rich-resource languages, multi-task learning, and knowledge distillation to achieve high speech quality and recognition accuracy even with minimal training data. This paves the way to broader language coverage and caters to the needs of diverse communities,  especially those with low-resource languages. 

\subsection{Language models for Kinyarwanda and Kiswahili}

In the paper \cite{elamin2023multilingual}, the authors developed a multilingual Automatic Speech Recognition (ASR) model for Kinyarwanda, Swahili, and Luganda. They utilized the Common Voice project's African language datasets and fine-tuned a pre-trained Conformer model with Connectionist Temporal Classification (CTC) decoding and Byte Pair Encoding (BPE) tokenization. The results demonstrated that the Kinyarwanda model achieved a WER of 17.57, while the average WER across all three languages was 21.91.

The authors \cite{rutunda2023kinyarwanda} developed a text-to-speech (TTS) model for Kinyarwanda, a Bantu language spoken in Rwanda, by leveraging an existing Kinyarwanda speech-to-text (STT) model and aligning audio recordings of the Kinyarwanda Bible with their corresponding text using CTC-Segmentation23. This resulted in a dataset containing 67.84 hours of studio-quality audio from multiple speakers. The TTS model was trained using the YourTTS framework, which supports multilingual and multi-speaker capabilities. The synthesized audio achieved a WER of 30.09\% compared to natural speech. Native Kinyarwanda speakers rated the naturalness of the synthesized speech with an average Mean Opinion Score (MOS) of 2.3 (on a scale of 1 to 5, where 5 represents natural human speech), and approximately 87\% of the synthesized samples were rated as intelligible or partially intelligible by listeners.

The paper \cite{rono2022development} describes the development of a Kiswahili TTS system using Tacotron 2 architecture and WaveNet vocoder. Tacotron 2 is a sequence-to-sequence model that consists of an encoder, a decoder, and a vocoder. The encoder converts input text into a sequence of characters, the decoder predicts Mel-spectrograms for each character sequence, and the vocoder transforms Mel-spectrograms into speech waveforms. The Kiswahili TTS system achieved a Mean Opinion Score (MOS) of 4.05, indicating that the generated speech is comparable to human speech.

Despite the abundance of models discussed in existing literature, a critical gap remains: most of these models lack flexibility in customization for edge devices. Consequently, deploying them on small edge devices without a back-up of high-end servers through a network remains a major challenge. Another limitation of the aforementioned approaches is that they fail to address scenarios where edge devices lack a consistently reliable Internet connection. To enhance computational efficiency and minimize performance compromises, it is essential to develop models optimized for seamless execution on edge devices. These models should be compressed to perform well locally and only offload processing to the Cloud when local computation cannot satisfy a desired performance level.

%% file: Sections/Approach.tex
\section{\MakeUppercase{Approach}}\label{sec:approach}

\begin{figure}[h]
    \centering
    \includegraphics[width=.99\linewidth]{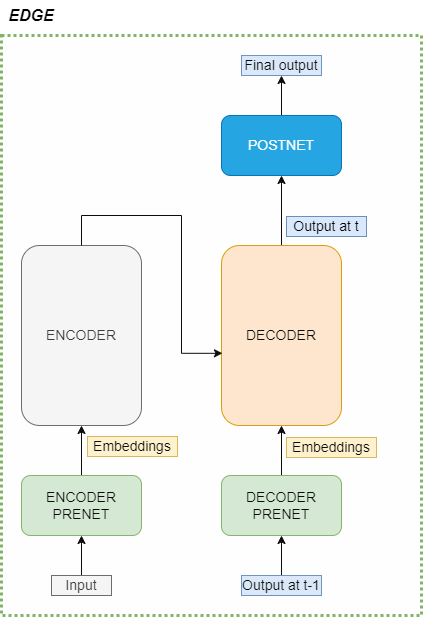}
    \caption{A generalized transformer architecture. The inputs are speech and text data for STT and TTS respectively. The outputs are text and speech for STT and TTS respectively. For both tasks, the decoder's inference is autoregressive }
    \label{fig:architecture_stt}
\end{figure}

In our approach, we adopt a Transformer-based architecture for both TTS and STT tasks, recognizing its widespread usage and effectiveness. Transformers employ a sequence-to-sequence model architecture characterized by multihead attention mechanisms, Encoder-Decoder modules, embedding layers, and both Prenet and Postnet components \cite{Conformer}, \cite{Guo},\cite{vaswani2023attentionneed}. In the STT process, the encoder receives audio input and the decoder produces text autoregressively. Conversely, in the TTS process, the encoder processes text input, and the decoder generates speech signals autoregressively, as illustrated in Figure \ref{fig:architecture_stt}.

Initially, the raw input data is processed by the encoder Prenet, which transforms it into suitable embeddings for the Encoder. The Encoder then generates contextualized embeddings, capturing the essential information from the input sequence. These embeddings are subsequently fed into the Decoder, which also incorporates embeddings from the Decoder Prenet. Operating in an autoregressive manner, the Decoder utilizes the output from the previous time step (Output at t-1) to generate the current output (Output at t). Finally, the sequence is processed through the Postnet, to get the final output. This systematic approach leverages the cross-attention mechanism within the Transformer architecture, enabling a context-aware sequence generation. Based on experience, such models tend to be large and resource-intensive, making them challenging to deploy on small-edge devices.

\begin{figure*}[h!]
    \centering
    \includegraphics[width=.99\linewidth]{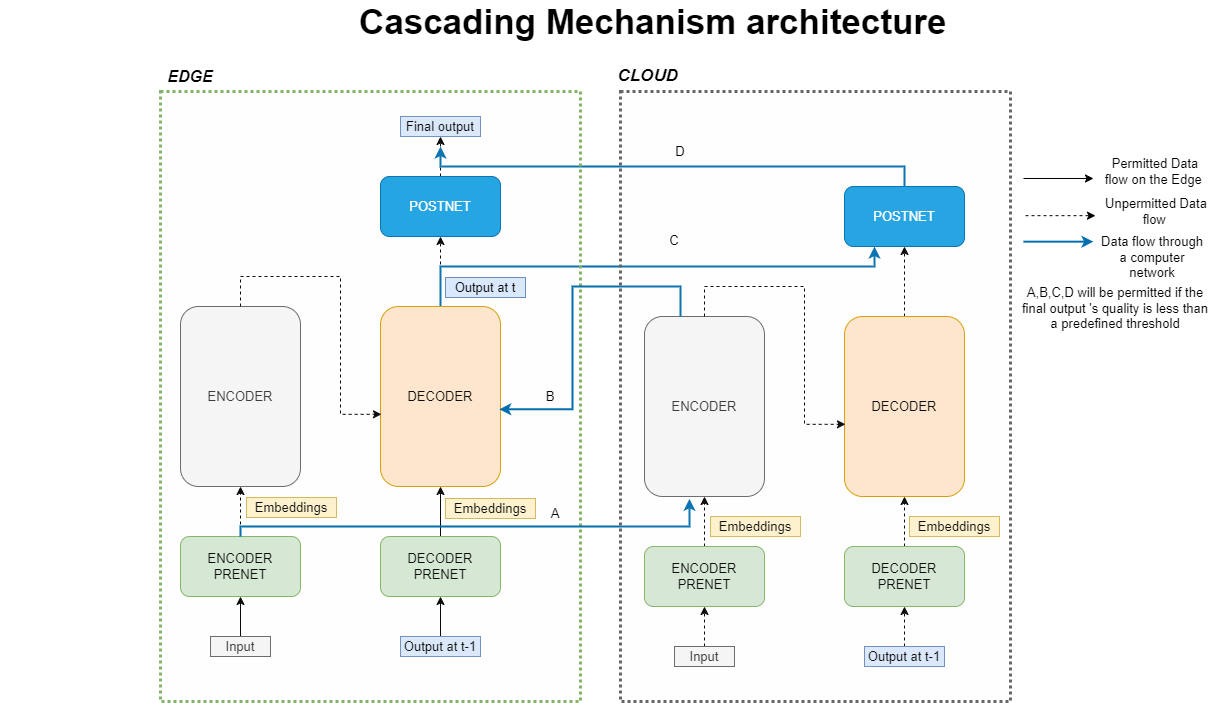}
    \caption{The cascading architecture is presented. On the left, the edge-level model is used to produce output. When the output's quality does not meet a predefined threshold, the internal data representations will be sent to the cloud for enhanced processing.}
    \label{fig:architecture_cascade}
\end{figure*}

Our work, as depicted in Figure \ref{fig:architecture_cascade}, proposed a cascading architecture for both the STT and the TTS tasks. These models comprise two interconnected components: an edge encoder-decoder structure and a cloud encoder-decoder structure. The edge encoder processes incoming data (audio signals for STT and text for TTS) and locally extracts relevant features. If the quality of the final output falls below a predefined threshold, the Cloud architecture is utilized. Optimized for low-latency inference, the edge encoder operates efficiently on resource-constrained devices. The Cloud encoder, located remotely, receives feature embeddings from the edge encoder when necessary. The final output from the Cloud encoder is then transmitted back to the edge device. The decoder in the cloud is skipped due to its use of autoregression in inference, which would increase network usage, thus leading to latency.

\subsection{Speech-To-Text based on Whisper model}

\begin{figure}[h]
    \centering
    \includegraphics[width=.99\linewidth]{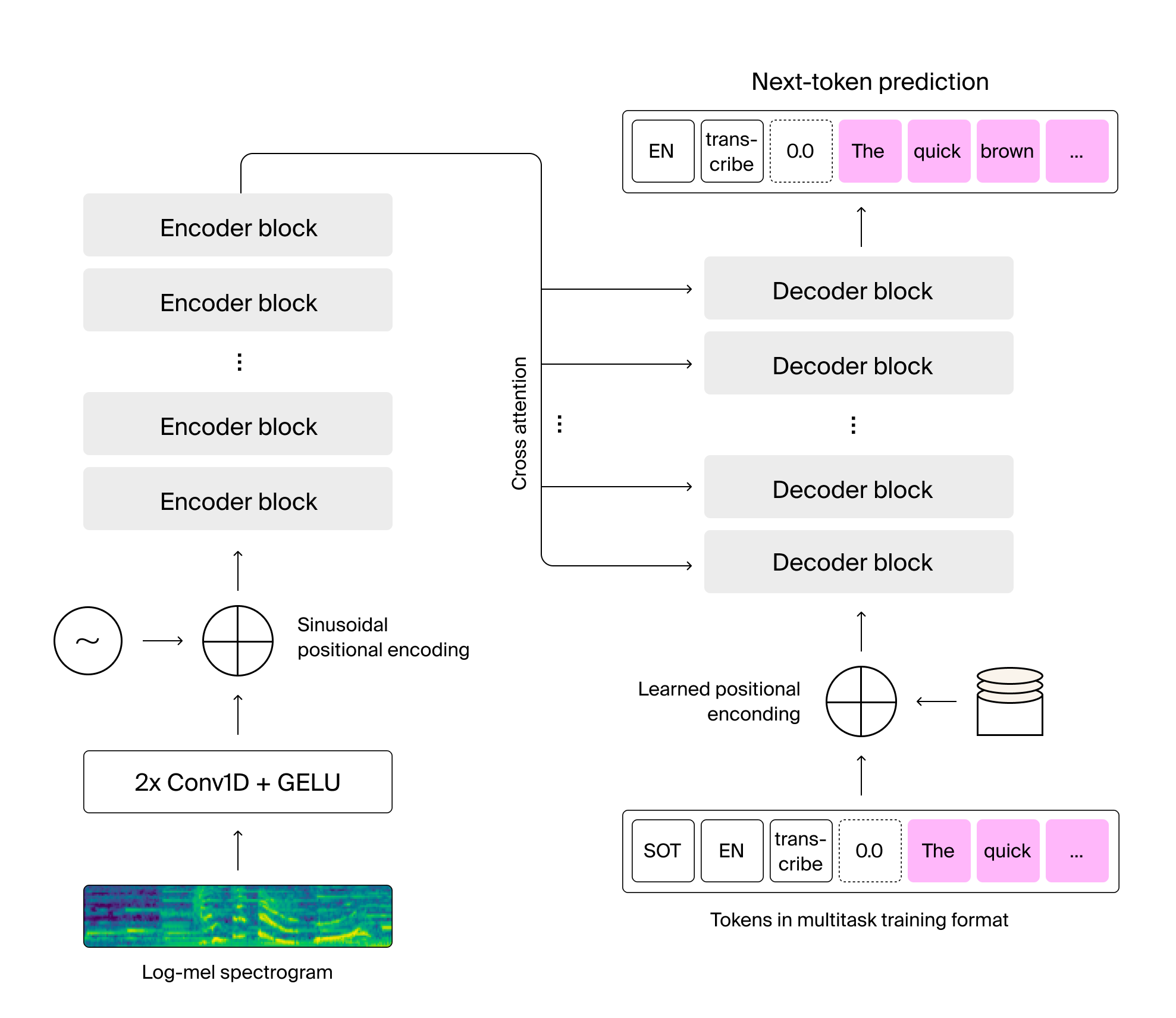}
    \caption{The Whisper architecture a simple end-to-end approach, implemented as an encoder-decoder Transformer \cite{Alec}.}
    \label{fig:whisper}
\end{figure}

In these experiments, we explored the development of a speech-to-text synthesis system for Kinyarwanda and Swahili languages. We leveraged the power of a pre-trained model called Whisper (shown in Figure \ref{fig:whisper}) and fine-tuned it on Mozilla Common Voice datasets specific to the Kinyarwanda and Swahili languages \cite{commonvoice:2020}.
Whisper, as it was introduced in \cite{Alec}, is a transformer-based neural network architecture designed for a variety of speech processing tasks, including automatic speech recognition (ASR) and speech translation. The main purpose of this model was to create an ASR model that ‘works reliably without the need for dataset-specific fine-tuning to achieve high-quality results on specific distributions’. It was trained on 680,000 hours of multilingual and multitasking-supervised data collected from the web. The Whisper architecture is based on an encoder-decoder transformer. This model enabled transcription in multiple languages as well as translation from those languages into English.

Whisper's end-to-end design converts audio input into a log-Mel spectrogram, which is processed by the model to produce text transcripts and additional information, simplifying traditional speech processing systems. The model's training on a diverse, multilingual dataset enables it to handle various speech characteristics and perform multiple tasks, including speech recognition and language identification. These characteristics made Whisper an ideal pre-trained model for our text-to-speech synthesis project in Kinyarwanda and Swahili. By fine-tuning Whisper on language-specific datasets, we specialized its capabilities for high-quality speech generation in the two targeted languages.

\subsection{Text-To-Speech, using SpeechT5 model}

For the TTS case, we explored the usage of the SpeechT5 alternative for Kinyarwanda and Swahili language speech synthesis. Here, we leveraged the capabilities of a pre-trained model called SpeechT5, fine-tuned on the Mozilla Common Voice datasets for these languages.

\begin{figure}[h]
    \centering
    \includegraphics[width=.99\linewidth]{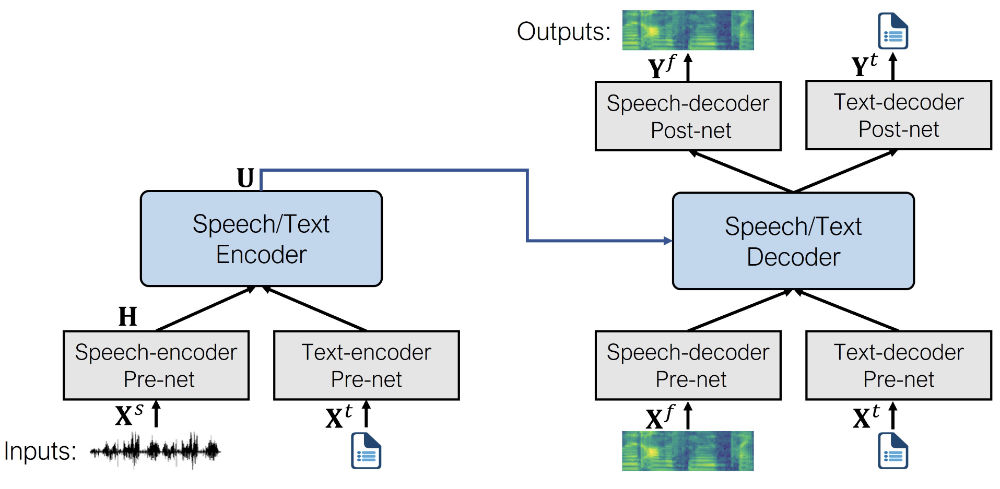}
    \caption{The model architecture of SpeechT5, which contains an encoder-decoder module and six modal specific pre/post-nets\cite{Junyi}.}
    \label{fig:speechT5}
\end{figure}

As Figure \ref{fig:speechT5} shows, SpeechT5 \cite{Junyi}, is a unified pre-trained model designed for various spoken language processing tasks. This model builds upon the success of T5 (Text-to-Text Transfer Transformer) but extends its capabilities to the speech domain. Below is a detailed description of SpeechT5's key features:

SpeechT5's unified encoder-decoder architecture processes both speech and text, enabling tasks like text-to-speech synthesis through a shared representation of language. Pre-trained on extensive speech and text data, SpeechT5 has capabilities for understanding linguistic structures and generating natural speech in multiple languages. Its speech synthesis capabilities also allow effective tuning on specific datasets which improve performance in languages such as Kinyarwanda and Swahili.

\subsection{Data collection}

For both TTS and STT, we used the Mozilla Common Voice dataset \cite{commonvoice:2020}, a diverse collection of community-supported Kinyarwanda and Kiswahili voice recordings and their transcription publicly available and widely recognized in speech technology research. We selected a subset dataset that comprised a total sample of 256 voice recordings, along with their corresponding textual transcriptions, for both Kiswahili and Kinyarwanda languages. To ensure the robustness and generalizability of our models, we divided the dataset into two distinct subsets: training and testing. The training set, which constitutes the majority of the data, is used to train our models, allowing them to learn the patterns and nuances of Kinyarwanda and Kiswahili speech. This process involves exposing the models to a wide array of vocal tones, dialects, and speech contexts, thereby enhancing their ability to accurately transcribe and synthesize speech in these languages. The testing set, on the other hand, serves a critical role in evaluating the performance of our models.

\subsection{Cascading Architecture}

To enhance the accuracy of our system, we implemented a cascading mechanism in both the STT and the TTS processes. For STT, we utilized the Whisper model at the edge. The model's final output logits were averaged and compared to a predefined threshold. If the average confidence level fell below this threshold, the cascading mechanism was triggered. This mechanism involved sending the internal representation of the audio data to the cloud for further processing, thereby leveraging the cloud's more robust computational resources to achieve higher accuracy.

For TTS, the system evaluated the final output using the signal-to-noise ratio (SNR) metric. If the SNR was found to be low, indicating potential noise in the generated speech, the cascading mechanism was again activated. The internal representation of the text was transmitted to the cloud, where advanced algorithms could refine the output. Averaging logits and detecting low signal-to-noise ratios are efficient techniques. By incorporating these cascading mechanisms, we significantly reduced the edge computing requirements for both STT and TTS processes. Edge inference shows CPU times of up to 8 seconds for long sequence speech-to-text, under 4 seconds for text-to-speech, and under 3 seconds for short sequence tasks on a 1.7 GHz CPU.

\section{Model Training}\label{Model Training}

\subsection {Speech To Text}

\subsubsection{Whisper ASR model training}
The Whisper ASR model is based on a standard Transformer-based encoder-decoder architecture. It follows the architecture of a log-Mel spectrogram input to the encoder, with cross-attention mechanisms connecting the encoder and decoder. The decoder autoregressively predicts text tokens, conditioned on the encoder’s hidden states and previously predicted tokens. The model comes in various configurations, including tiny, base, small, medium, and large. Each configuration has different numbers of layers, widths, and heads. In this paper, we fine-tuned the multilingual version of the “small” checkpoint with approximately 244 million parameters.

During fine-tuning, we employed several hyperparameters and a structured training process. The learning rate was set to 1e-5 to allow for gradual weight updates during training, and warmup\_steps were used to gradually increase the learning rate during the initial stages. The training process was defined by a set number of max\_steps and included gradient\_accumulation\_steps to accumulate gradients, simulating larger batch sizes for efficiency. We enabled mixed-precision training (fp16) for faster convergence. The evaluation strategy involved assessing the model every 1000 steps, with the generation\_max\_length set to 225 to define the maximum output sequence length during inference. Model checkpoints were saved every 1000 steps, and the model was evaluated on the test set at the same interval. Training progress was logged every 25 steps to keep track of the model's performance. The metric for selecting the best model was based on the word error rate (WER), and the best model was loaded at the end of the training process. Finally, we enabled push\_to\_hub, allowing the trained model to be shared and version-controlled on the Hugging Face Hub.

In our approach, the following parameters were used:

\begin{lstlisting}[language=Octave,caption={Model training parameters},basicstyle=\scriptsize,label={lst:parameter}]
train_args={
    per_device_train_batch_size="16",
    gradient_accumulation_steps="2",
    learning_rate="1e-5",
    warmup_steps="500",
    max_steps="4000",
    gradient_checkpointing="True",
    fp16="True",
    evaluation_strategy="steps",
    per_device_eval_batch_size="8",
    save_steps="1000",
    eval_steps="1000",
    logging_steps="25",
    load_best_model_at_end="True",
    metric_for_best_model="wer",
    greater_is_better="False",
    label_names=["labels"],
    loss: "CrossEntropy"
    push_to_hub=True
}
\end{lstlisting}

\subsubsection{Evaluation Metric}
To assess the performance of our fine-tuned model, we used the WER metric. WER calculates the number of errors (insertions, substitutions, and deletions) an automatic speech recognition system makes when compared to a human-generated reference transcript.  In the context of speech-to-text transcription, a lower WER indicates a higher fidelity between the synthesized speech and the original written text.

\subsection {Text To Speech}
\subsubsection{SpeechT5 Model training}
We fine-tuned the Speech-T5 model using the \textit{Seq2SeqTrainer} from the Transformers library. The process began with tokenizing the input text, and converting it into a sequence of token IDs. These token IDs, along with speaker embeddings, were then fed into the encoder. The encoder processed these inputs to generate hidden state representations. The decoder was subsequently trained to predict the log-Mel spectrogram from these hidden states, effectively learning to map the text and speaker information to the corresponding speech features. The log-Mel spectrogram would then be converted to audio using Microsoft's SpeechT5HifiGan. This end-to-end training approach allowed the model to generate speech outputs from textual inputs.

In our approach, the following parameters were used:
\begin{lstlisting}[language=Octave,caption={Model training parameters},basicstyle=\scriptsize,label={lst:parameters}]
train_args={
    per_device_train_batch_size="16",
    gradient_accumulation_steps="1",
    learning_rate="1e-5",
    warmup_steps="500",
    max_steps="4000",
    gradient_checkpointing="True",
    fp16="True",
    evaluation_strategy="steps",
    per_device_eval_batch_size="8",
    predict_with_generate="True",
    generation_max_length="225",
    save_steps="1000",
    eval_steps="1000",
    logging_steps="25",
    load_best_model_at_end="True",
    greater_is_better="False",
    push_to_hub="True"
}
\end{lstlisting}

This configuration batch size for training is set to 16 examples, and gradients are accumulated over 2 steps before updating the model to improve efficiency. The learning rate is established at 1e-5, with 500 warmup steps to gradually increase it. Training will conclude after 4000 steps. Gradient checkpointing is enabled to save memory, and mixed precision (FP16) training is utilized for faster computations.

The model was evaluated every 1000 steps, with an evaluation batch size of 8 examples. Model checkpoints are also saved every 1000 steps. Training progress is logged every 25 steps, with logs reported to TensorBoard \footnote{\href{https://www.tensorflow.org/tensorboard}{https://www.tensorflow.org/tensorboard}} for better tracking. At the end of training, the best model based on validation performance will be loaded. The target label name is specified as “labels” for the dataset.

\subsubsection{Evaluation Metrics}
In this text-to-speech exploration, we employed mean-square error loss (MSE loss) as a performance indicator. During model training, the model is exposed to both training and validation data. The training data is used to adjust the model's internal parameters, while the validation data helps assess how well the model generalizes to unseen examples. Validation loss measures the model's performance on the validation data. Lower validation loss signifies better model performance during fine-tuning, indicating the model is learning to generate speech that aligns well with the ground truth (written text).

%% file: Sections/results.tex
\section{\MakeUppercase{Experiments and Results }}\label{sec:results}

\begin{table*}
    \centering
    \begin{tabular}{|c|c|c|c||c|c|c|c|}
        \hline
        \multicolumn{4}{|c||}{Swahili Fine-tuned Whisper Model Results} & \multicolumn{4}{c|}{Kinyarwanda Fine-tuned Whisper Model Results} \\ \hline
        \textbf{Step} & \textbf{Training Loss} & \textbf{Validation Loss} & \textbf{WER} & \textbf{Step} & \textbf{Training Loss} & \textbf{Validation Loss} & \textbf{WER} \\ \hline
        20  & 1.7887 & 2.1523 & 26.3736 & 20  & 2.1422 & 2.3111 & 34.5953 \\ \hline
        40  & 1.7887 & 2.0751 & 26.2515 & 40  & 2.1422 & 2.2022 & 33.9426 \\ \hline
        60  & 1.6873 & 2.0161 & 26.3736 & 60  & 1.8742 & 2.1406 & 33.6815 \\ \hline
        80  & 1.5626 & 1.9788 & 26.3736 & 80  & 1.7608 & 2.1077 & 33.2898 \\ \hline
        100 & 1.4991 & 1.9641 & 26.3736 & 100 & 1.6573 & 2.0954 & 33.4204 \\ \hline
    \end{tabular}
    \caption{ { Results after Fine-tuning the Whisper ASR Model on Swahili and Kinyarwanda Datasets: Training and validation losses are measured using cross-entropy loss. Word Error Rate (WER) is given as a percentage, indicating errors (insertions, substitutions, deletions) compared to a human-generated reference transcript. }}
    \label{tab:whisper_results}
\end{table*}

\begin{table*}
    \centering
    \begin{tabular}{|c|c|c||c|c|c|}
        \hline
        \multicolumn{3}{|c||}{Swahili Fine-tuned SpeechT5 Model Results} & \multicolumn{3}{c|}{Kinyarwanda Fine-tuned SpeechT5 Model Results} \\ \hline
        \textbf{Step} & \textbf{Training Loss} & \textbf{Validation Loss} & \textbf{Step} & \textbf{Training Loss(} & \textbf{Validation Loss} \\ \hline
        1000 & 0.598900 & 0.553199 & 1000 & 0.695200 & 0.991951 \\ \hline
        2000 & 0.564900 & 0.534779 & 2000 & 0.477100 & 0.926026 \\ \hline
        3000 & 0.562600 & 0.526816 & 3000 & 0.313200 & 0.950552 \\ \hline
        4000 & 0.556600 & 0.523976 & 4000 & 0.533400 & 0.497876 \\ \hline
    \end{tabular}
     \caption{ Results after Fine-tuning the SpeechT5 Text-to-Speech Model on Swahili and Kinyarwanda Datasets: Training and validation losses are measured using mean squared error (MSE) loss.}
    \label{tab:speecht5_results}
\end{table*}

\subsection{Speech-to-Text:}
The Whisper model was fine-tuned for each language independently, where the checkpoints in \cite{dmusingu} and \cite{mbazaWhisper} were used for Swahili and Kinyarwanda, repsectively. The process achieved a WER of 26.37\% for Swahili and 33.42\% for Kinyarwanda. The training process did not significantly change the initial WER. However, the improving trends of  the cross-entropy loss (shown in in Table \ref{tab:whisper_results})  promise improved performances, given additional model finetuning. 

\subsection{Text-to-Speech}

Table \ref{tab:speecht5_results} presents the results of fine-tuning the SpeechT5 text-to-speech model on Swahili and Kinyarwanda datasets and reveal distinct trends in training and validation MSE losses, reflecting the model's adaptability and learning efficiency across different languages. For Swahili, the training loss shows a consistent decrease from 0.59 at step 1000 to 0.56 at step 4000, accompanied by a parallel reduction in validation loss from 0.55 to 0.52. This steady improvement across both metrics indicates a robust enhancement in the model's performance, suggesting effective learning and generalization on the Swahili dataset.

In contrast, the Kinyarwanda results display more fluctuating patterns in the MSE losses. The training loss significantly drops from 0.69 at step 1000 to 0.31 at step 3000, only to rise again to 0.53 at step 4000. The validation loss mirrors this volatility, starting at 0.991951, decreasing to 0.93, rising slightly, and then dropping sharply to 0.49 by step 4000. This erratic behavior suggests initial challenges in model generalization on the Kinyarwanda dataset, but ultimately, the substantial decrease in validation loss by the final step indicates significant progress, highlighting the model's capacity to adjust and improve through the training process.

%% file: Sections/Deployment_On_Edge.tex
\section{Deployment On The Edge}\label{sec:deployment}

This study presents a cascading architecture that utilizes both edge and Cloud resources for the deployment of Text-to-Speech (TTS) and Speech-to-Text (STT) models. The architecture incorporates two distinct models for each task: a compressed encoder model deployed at the edge and a non-compressed encoder model hosted in the Cloud. As depicted in Figure \ref{fig:architecture_cascade}, the transformer model is deployed in a cascaded manner, allowing engineered features from the edge model’s encoder pre-net layer to be transmitted to the Cloud model, and enhanced hidden states to be retrieved from the Cloud back to the edge.

On a computer equipped with 16 GB of memory and a 1.7 GHz CPU, NetLimiter \cite{netlimiter} was employed to simulate various local network bandwidths. This setup was used to compare the inference times required to generate speech for both short and long texts using the finetuned SpeechT5 model. The simulation involved a simple Swahili greeting, “Habari gani?” (literally translating to “How is it going?”), consisting of 12 characters, and the first paragraph from \cite{kenya_wikipedia} translated into Swahili, resulting in a text of 270 characters.

Swahili text: \textit{“Hali ya hewa ya Kenya inatofautiana kutoka kitropiki kando ya pwani hadi nchi kavu hadi sehemu ya kaskazini na kaskazini mashariki mwa nchi. Eneo hilo hupokea mwanga mwingi wa jua kila mwezi. Kawaida ni baridi usiku na mapema asubuhi ndani ya nchi kwenye miinuko ya juu.”}

Translation: \textit{ “Kenya's climate varies from tropical along the coast to temperate inland to arid in the north and northeast parts of the country. The area receives a great deal of sunshine every month. It is usually cool at night and early in the morning inland at higher elevations.” (copied from \cite{kenya_wikipedia})}

\begin{figure}[h]
    \centering
    \includegraphics[width=.99\linewidth]{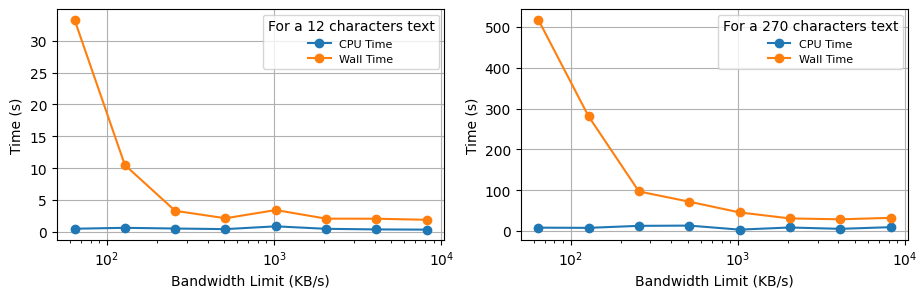}
    \caption{Simulation results of a 12-characters and a 270-characters Swahili text’s speech synthesis time against different bandwidth values. The CPU time is the time taken by the edge device in computation while the wall time is the total elapsed time.}
    \label{fig:figure1_BW}
\end{figure}

As illustrated in Figure \ref{fig:figure1_BW}, it is evident that the longer sequence required more processing time. Additionally, the results indicate a very significant increase in inference time when the bandwidth approaches or falls below 512 KB/s. For bandwidths exceeding 512 KB/s, generating a 1-second audio from a 12-character text took approximately 2 to 3 seconds. However, for a 270-character text, which produced a 19-second audio, the inference time was around 30 seconds for bandwidths greater than 2 MB/s, but it began to increase as the bandwidth decreased. Therefore, this cascaded framework demonstrates time efficiency with smaller bandwidths, provided that the input texts are short.

\begin{figure}[h]
    \centering
    \includegraphics[width=.99\linewidth]{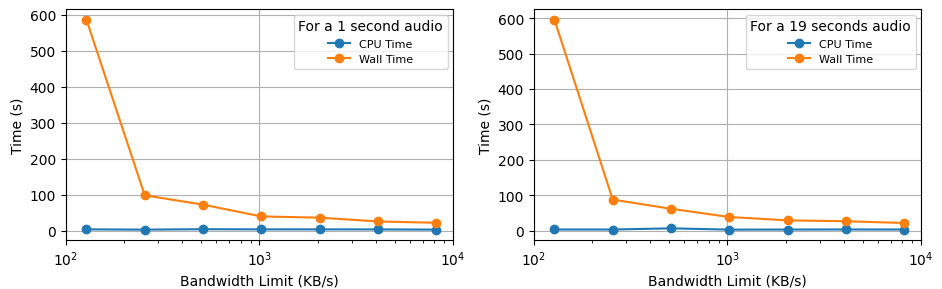}
    \caption{Simulation results of a 1-second and a 19-seconds Swahili audio transcription time against different bandwidth values. The CPU time is the time taken by the edge device in computation while the wall time is the total elapsed time.}
    \label{fig:figure2_BW}
\end{figure}

\begin{figure}[h]
    \centering
    \includegraphics[width=.99\linewidth]{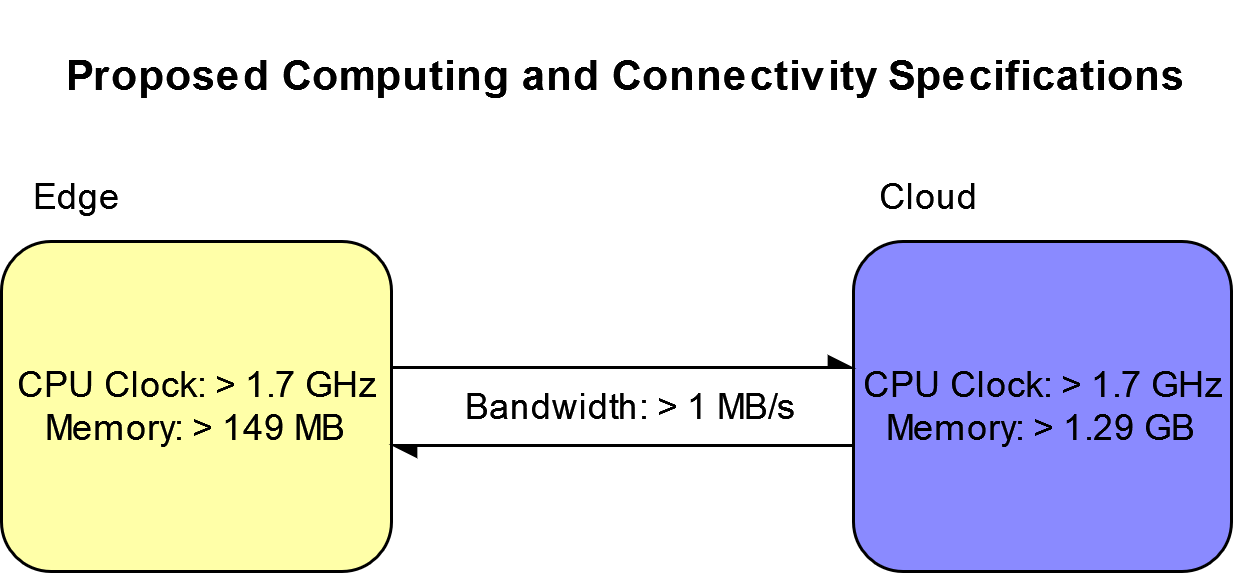}
    \caption{Proposed minimum computing and connectivity requirements for the edge device and the cloud 
}
    \label{fig:cascading_illustration}
\end{figure}

For text transcription, the same method was applied to simulate various bandwidths, using the generated speech audios as input and the finetuned Whisper model within the cascaded framework. Figure \ref{fig:figure2_BW} illustrates the inference times for a 1-second “Habari Gani?” audio and a 19-second audio derived from a 270-character text, in relation to bandwidth limits. As shown in Figure \ref{fig:figure2_BW}, the inference time generally remains constant regardless of sequence length, due to the fixed length of the model’s encoder output, which requires the same duration to transmit back to the edge model. However, both graphs indicate a significant increase in inference time when the bandwidth drops below 1024 KB/s. These results suggest that both short and long audio files will take nearly the same time to process. Consequently, a minimum bandwidth of 1024 KB/s (1 MB/s) is recommended for deploying this framework, as detailed in Figure \ref{fig:cascading_illustration}.

Using Kenya as a case study, the 2023-2024 bandwidth data from \cite{kenya_isps} indicate that the majority of Internet Service Providers (ISPs) offer download speeds exceeding 1 MB/s (equivalent to 8 Mb/s, as per the dataset). This suggests that, on average, the optimal bandwidths required for the proposed cascaded framework (i.e., bandwidths of at least 512 KB/s for TTS and 1 MB/s for STT) are accessible throughout the country.

In addition to bandwidth considerations, the computing capabilities of the edge devices are also very important. Monitoring the models’ sizes, in terms of parameters and data types, reveals that the cascaded framework utilizes 38\% of the SpeechT5 model and 56\% of the Whisper model on the edge’s memory. The edge device must manage a load of 226 MB for the SpeechT5 model and 567 MB for the Whisper model, which can be reduced by up to 25\% through quantizing the model parameters from FP32 to INT8 data types \cite{ahn2023performance}. Consequently, the cascaded framework theoretically requires edge devices to provide at least 57 MB (25\% of 226 MB) of memory for the SpeechT5 model and at least 149 MB (25\% of 567 MB) for the Whisper model. This translates to an overall edge memory usage of 9.5\% for the SpeechT5 model and 14\% for the Whisper model which seems quite reasonable.

\begin{figure}[h]
    \centering
    \includegraphics[width=.99\linewidth]{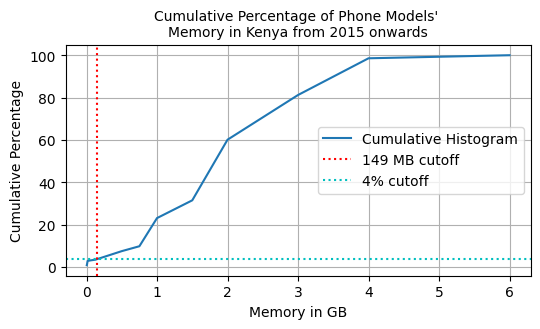}
    \caption{Cumulative percentage of phone models' memory in Kenya. Note that only 4\% can't meet the requirement of the proposed cascaded architecture.
}
    \label{fig:cumulative_memory}
\end{figure}
Based on the market share data of different brands of mobile phones in Kenya \cite{statista2024} and the phone models’ specifications dataset \cite{gsmarena2024}, the cumulative percentage of phone models’ memory in Kenya was calculated. As illustrated in Figure \ref{fig:cumulative_memory}, less than 4\% of phone models do not meet the required 149 MB of memory for deploying the proposed framework. This indicates that over 96\% of phone models in Kenya are suitable for utilizing the benefits of the cascaded architecture presented in this paper.


Considering the computation speed of edge devices is important as it highlights the performance of devices used in the East African region. The experiments in this study utilized an edge device with a 1.7 GHz CPU clock rate. Given that the model inference on the edge is not parallelized, the CPU time for any other edge device can be estimated, as it is inversely proportional to the clock rate. Using the market data on mobile phones in Kenya \cite{statista2024} and the phone models’ specifications dataset \cite{gsmarena2024}, CPU time approximations were calculated based on the average CPU times observed on the 1.7 GHz CPU device.

\begin{figure}[h]
    \centering
    \includegraphics[width=.99\linewidth]{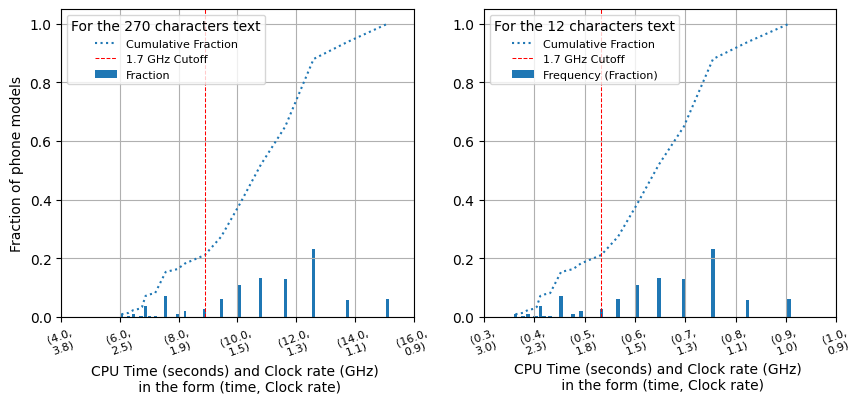}
    \caption{Fraction of phone models that can synthesize speech within the CPU time required. (On the left: graph for the 270 characters text, On the right: graph for the 12 characters text)
.}
    \label{fig:figure1}
\end{figure}

\begin{figure}[h]
    \centering
    \includegraphics[width=.99\linewidth]{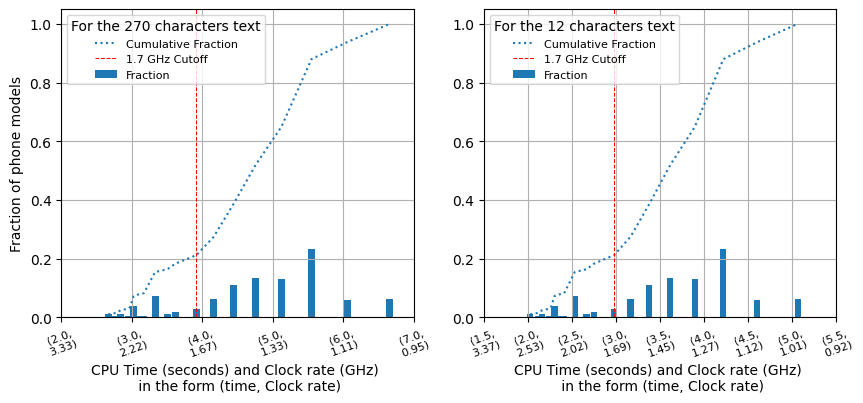}
    \caption{Fraction of phone models that can transcribe speech within the CPU time required. On the left: graph for the 270 characters text (from a 19 seconds audio), On the right: graph for the 12 characters text (from a 1 second audio)}
    \label{fig:figure2}
\end{figure}

For TTS CPU time, it was observed that the 1.7 GHz CPU device took an average of 0.5 seconds for a 12-character text and 8.9 seconds for a 270-character text. Figure \ref{fig:figure1}  indicates that approximately 20\% of phone models could surpass this performance. However, Figure \ref{fig:figure1} also demonstrates that a 1 GHz CPU device’s CPU time for the 12-character text remains under 1 second, which is acceptable.

For STT CPU time, it was observed that the 1.7 GHz CPU device took an average of 2.97 seconds for a 12-character text and 3.92 seconds for a 270-character text. Figure \ref{fig:figure2} indicates that approximately 20\% of phone models could surpass this performance. However, Figure \ref{fig:figure2} also shows that increasing the input text from 12 characters to 270 characters only increases the edge’s CPU time by about 0.9 seconds, demonstrating the proposed architecture’s capability to accommodate low-clock-rate CPUs.

\begin{figure}[h]
    \centering
    \includegraphics[width=.99\linewidth]{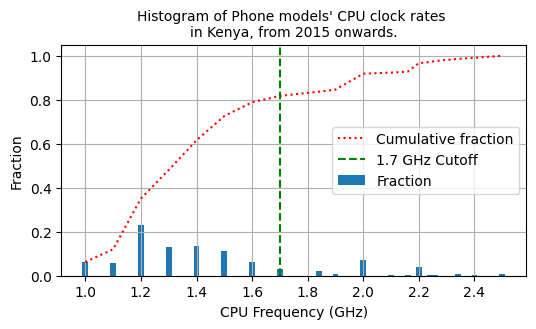}
    \caption{Histogram of CPU clock rate of phone models in Kenya.
}
    \label{fig:cpu_histogram}
\end{figure}

Figure \ref{fig:cpu_histogram} indicates that approximately 80\% of phone models used in East Africa have a CPU clock rate below 1.7 GHz. This suggests that the time required for TTS and STT translation will be longer than the previously mentioned values. This further motivates and supports the need for the cascaded architecture proposed in this paper. Additionally, messages such as news items on popular TV channels, other media outlets, or financial market updates could easily exceed 270 characters, implying that the actual translation times for TTS and STT might be significantly longer. By utilizing the cascaded architecture and the input sequence length versus inference time insights provided in this paper, the waiting time for TTS and STT could be reduced to acceptable levels, thus providing a good end-user experience.

%% file: Sections/Conclusion.tex
\section{\MakeUppercase{Conclusion and Future Work}}\label{sec:futurework}

The paper presents a novel approach for edge-based speech transcription and synthesis targeting the Kinyarwanda and Swahili languages, leveraging a hybrid edge-cloud network architecture. This system processes speech data locally on edge devices and escalates to cloud-based resources when necessary, facilitating efficient use of computational resources and enhanced accessibility. Key advantages of this approach include the efficient handling of local language processing on low-resource devices which are typical in Eastern Africa, enabling real-time speech processing, and reducing reliance on cloud compute services. By integrating local processing capabilities with cloud-based support, the system ensures that speech data can be managed efficiently, even in regions with limited access to high-end computational infrastructure. This hybrid approach not only optimizes resource usage but also enhances the accessibility and reliability of speech processing technologies in underrepresented languages, such as Kinyarwanda and Swahili.

Future research should explore optimizing the models used in this paper further to achieve more accurate and efficient transcription results, potentially expanding their applicability to other low-resource languages and diverse speech environments. In addition, a practical evaluation of the computational overhead and resource requirements of the proposed hybrid network architecture could provide valuable insights into its scalability and effectiveness in various deployment scenarios.

%% file: references.bib
@incollection{waliaula2023,
  title={Kiswahili in Kenya: Broken Language and Broken Promises},
  author={Waliaula, Ken Walibora},
  booktitle={The Palgrave Handbook of Contemporary Kenya},
  pages={267--277},
  year={2023},
  publisher={Springer}
}

@misc{RwandaPop,
  author       = "{National Institute of Statistics Rwanda}",
  title        = "{Population size and Population characteristics}",
  howpublished = "\url{https://www.statistics.gov.rw/statistical-publications/subject/population-size-and-population-characteristics}",
  note         = "[Online; accessed 10-August-2024]"
}

@article{lisanza2021swahili,
  author       = "V. Lisanza",
  title        = "{Swahili gaining popularity globally}",
  journal      = "Africa Renewal",
  year         = "2021",
  month        = "Dec. 09",
  url          = {https://www.un.org/africarenewal/magazine/december-2021/swahili-gaining-popularity-globally},
  note         = "[Online; accessed 10-August-2024]"
}

@article{Metalom_Fendji_Yenke_2022b, title={Towards to a Direct Speech to Speech for Endangered Languages in Africa}, url={https://www.researchgate.net/publication/361984238_Towards_to_a_Direct_Speech_to_Speech_for_Endangered_Languages_in_Africa}, journal={ResearchGate}, author={Metalom, Diane Carole Carole Metalom Tala and Fendji, Jean Louis Kedieng Ebongue and Yenke, Blaise Omer}, year={2022}, month=oct }

@INPROCEEDINGS{Atienza,
  author={Atienza, Rowel},
  booktitle={ICASSP 2023 - 2023 IEEE International Conference on Acoustics, Speech and Signal Processing (ICASSP)}, 
  title={EfficientSpeech: An On-Device Text to Speech Model}, 
  year={2023},
  volume={},
  number={},
  pages={1-5},
  doi={10.1109/ICASSP49357.2023.10094639}}

@misc{wang2017synthesis,
      title={Tacotron: Towards End-to-End Speech Synthesis}, 
      author={Yuxuan Wang and RJ Skerry-Ryan and Daisy Stanton and Yonghui Wu and Ron J. Weiss and Navdeep Jaitly and Zongheng Yang and Ying Xiao and Zhifeng Chen and Samy Bengio and Quoc Le and Yannis Agiomyrgiannakis and Rob Clark and Rif A. Saurous},
      year={2017},
      eprint={1703.10135},
      archivePrefix={arXiv},
      primaryClass={cs.CL},
      url={https://arxiv.org/abs/1703.10135}, 
}

@INPROCEEDINGS{Shen,
  author={Shen, Jonathan and Pang, Ruoming and Weiss, Ron J. and Schuster, Mike and Jaitly, Navdeep and Yang, Zongheng and Chen, Zhifeng and Zhang, Yu and Wang, Yuxuan and Skerrv-Ryan, Rj and Saurous, Rif A. and Agiomvrgiannakis, Yannis and Wu, Yonghui},
  booktitle={2018 IEEE International Conference on Acoustics, Speech and Signal Processing (ICASSP)}, 
  title={Natural TTS Synthesis by Conditioning Wavenet on MEL Spectrogram Predictions}, 
  year={2018},
  volume={},
  number={},
  pages={4779-4783},
  doi={10.1109/ICASSP.2018.8461368}}

@misc{WaveGlow,
      title={WaveGlow: A Flow-based Generative Network for Speech Synthesis}, 
      author={Ryan Prenger and Rafael Valle and Bryan Catanzaro},
      year={2018},
      eprint={1811.00002},
      archivePrefix={arXiv},
      primaryClass={cs.SD},
      url={https://arxiv.org/abs/1811.00002}, 
}

@misc{Anjuli,
      title={Streaming End-to-end Speech Recognition For Mobile Devices}, 
      author={Yanzhang He and Tara N. Sainath and Rohit Prabhavalkar and Ian McGraw and Raziel Alvarez and Ding Zhao and David Rybach and Anjuli Kannan and Yonghui Wu and Ruoming Pang and Qiao Liang and Deepti Bhatia and Yuan Shangguan and Bo Li and Golan Pundak and Khe Chai Sim and Tom Bagby and Shuo-yiin Chang and Kanishka Rao and Alexander Gruenstein},
      year={2018},
      eprint={1811.06621},
      archivePrefix={arXiv},
      primaryClass={cs.CL},
      url={https://arxiv.org/abs/1811.06621}, 
}

@inproceedings{eichner00_icslp,
  author={Matthias Eichner and Matthias Wolff and Rüdiger Hoffmann},
  title={{A unified approach for speech synthesis and speech recognition using stochastic Markov graphs}},
  year=2000,
  booktitle={Proc. 6th International Conference on Spoken Language Processing (ICSLP 2000)},
  pages={vol. 1, 701-704},
  doi={10.21437/ICSLP.2000-174},
  issn={2958-1796}
}

@misc{Rohit,
      title={End-to-End Speech Recognition: A Survey}, 
      author={Rohit Prabhavalkar and Takaaki Hori and Tara N. Sainath and Ralf Schlüter and Shinji Watanabe},
      year={2023},
      eprint={2303.03329},
      archivePrefix={arXiv},
      primaryClass={eess.AS},
      url={https://arxiv.org/abs/2303.03329}, 
}

@misc{Conformer,
      title={Conformer: Convolution-augmented Transformer for Speech Recognition}, 
      author={Anmol Gulati and James Qin and Chung-Cheng Chiu and Niki Parmar and Yu Zhang and Jiahui Yu and Wei Han and Shibo Wang and Zhengdong Zhang and Yonghui Wu and Ruoming Pang},
      year={2020},
      eprint={2005.08100},
      archivePrefix={arXiv},
      primaryClass={eess.AS},
      url={https://arxiv.org/abs/2005.08100}, 
}

@misc{Guo,
      title={Recent Developments on ESPnet Toolkit Boosted by Conformer}, 
      author={Pengcheng Guo and Florian Boyer and Xuankai Chang and Tomoki Hayashi and Yosuke Higuchi and Hirofumi Inaguma and Naoyuki Kamo and Chenda Li and Daniel Garcia-Romero and Jiatong Shi and Jing Shi and Shinji Watanabe and Kun Wei and Wangyou Zhang and Yuekai Zhang},
      year={2020},
      eprint={2010.13956},
      archivePrefix={arXiv},
      primaryClass={eess.AS},
      url={https://arxiv.org/abs/2010.13956}, 
}

@misc{Gabriel,
      title={End-to-end ASR: from Supervised to Semi-Supervised Learning with Modern Architectures}, 
      author={Gabriel Synnaeve and Qiantong Xu and Jacob Kahn and Tatiana Likhomanenko and Edouard Grave and Vineel Pratap and Anuroop Sriram and Vitaliy Liptchinsky and Ronan Collobert},
      year={2020},
      eprint={1911.08460},
      archivePrefix={arXiv},
      primaryClass={cs.CL},
      url={https://arxiv.org/abs/1911.08460}, 
}

@misc{HuBERT,
      title={HuBERT: Self-Supervised Speech Representation Learning by Masked Prediction of Hidden Units}, 
      author={Wei-Ning Hsu and Benjamin Bolte and Yao-Hung Hubert Tsai and Kushal Lakhotia and Ruslan Salakhutdinov and Abdelrahman Mohamed},
      year={2021},
      eprint={2106.07447},
      archivePrefix={arXiv},
      primaryClass={cs.CL},
      url={https://arxiv.org/abs/2106.07447}, 
}

@misc{Ren,
      title={FastSpeech 2: Fast and High-Quality End-to-End Text to Speech}, 
      author={Yi Ren and Chenxu Hu and Xu Tan and Tao Qin and Sheng Zhao and Zhou Zhao and Tie-Yan Liu},
      year={2022},
      eprint={2006.04558},
      archivePrefix={arXiv},
      primaryClass={eess.AS},
      url={https://arxiv.org/abs/2006.04558}, 
}

@misc{Rowel,
      title={EfficientSpeech: An On-Device Text to Speech Model}, 
      author={Rowel Atienza},
      year={2023},
      eprint={2305.13905},
      archivePrefix={arXiv},
      primaryClass={eess.AS},
      url={https://arxiv.org/abs/2305.13905}, 
}

@misc{Wang,
      title={Tacotron: Towards End-to-End Speech Synthesis}, 
      author={Yuxuan Wang and RJ Skerry-Ryan and Daisy Stanton and Yonghui Wu and Ron J. Weiss and Navdeep Jaitly and Zongheng Yang and Ying Xiao and Zhifeng Chen and Samy Bengio and Quoc Le and Yannis Agiomyrgiannakis and Rob Clark and Rif A. Saurous},
      year={2017},
      eprint={1703.10135},
      archivePrefix={arXiv},
      primaryClass={cs.CL},
      url={https://arxiv.org/abs/1703.10135}, 
}

@misc{Jonathan,
      title={Natural TTS Synthesis by Conditioning WaveNet on Mel Spectrogram Predictions}, 
      author={Jonathan Shen and Ruoming Pang and Ron J. Weiss and Mike Schuster and Navdeep Jaitly and Zongheng Yang and Zhifeng Chen and Yu Zhang and Yuxuan Wang and RJ Skerry-Ryan and Rif A. Saurous and Yannis Agiomyrgiannakis and Yonghui Wu},
      year={2018},
      eprint={1712.05884},
      archivePrefix={arXiv},
      primaryClass={cs.CL},
      url={https://arxiv.org/abs/1712.05884}, 
}

@misc{Ryan,
      title={WaveGlow: A Flow-based Generative Network for Speech Synthesis}, 
      author={Ryan Prenger and Rafael Valle and Bryan Catanzaro},
      year={2018},
      eprint={1811.00002},
      archivePrefix={arXiv},
      primaryClass={cs.SD},
      url={https://arxiv.org/abs/1811.00002}, 
}

@misc{Yanzhang,
      title={Streaming End-to-end Speech Recognition For Mobile Devices}, 
      author={Yanzhang He and Tara N. Sainath and Rohit Prabhavalkar and Ian McGraw and Raziel Alvarez and Ding Zhao and David Rybach and Anjuli Kannan and Yonghui Wu and Ruoming Pang and Qiao Liang and Deepti Bhatia and Yuan Shangguan and Bo Li and Golan Pundak and Khe Chai Sim and Tom Bagby and Shuo-yiin Chang and Kanishka Rao and Alexander Gruenstein},
      year={2018},
      eprint={1811.06621},
      archivePrefix={arXiv},
      primaryClass={cs.CL},
      url={https://arxiv.org/abs/1811.06621}, 
}

@inproceedings{Chen_2023,
   title={LightGrad: Lightweight Diffusion Probabilistic Model for Text-to-Speech},
   url={http://dx.doi.org/10.1109/ICASSP49357.2023.10096710},
   DOI={10.1109/icassp49357.2023.10096710},
   booktitle={ICASSP 2023 - 2023 IEEE International Conference on Acoustics, Speech and Signal Processing (ICASSP)},
   publisher={IEEE},
   author={Chen, Jie and Song, Xingchen and Peng, Zhendong and Zhang, Binbin and Pan, Fuping and Wu, Zhiyong},
   year={2023},
   month=jun }

@misc{Jin,
      title={LRSpeech: Extremely Low-Resource Speech Synthesis and Recognition}, 
      author={Jin Xu and Xu Tan and Yi Ren and Tao Qin and Jian Li and Sheng Zhao and Tie-Yan Liu},
      year={2020},
      eprint={2008.03687},
      archivePrefix={arXiv},
      primaryClass={eess.AS},
      url={https://arxiv.org/abs/2008.03687}, 
}

@article{Asante_Imamura_2019b, title={Speech Recognition and Speech Synthesis Models for Micro Devices}, volume={27}, url={https://www.itm-conferences.org/articles/itmconf/abs/2019/04/itmconf_dictap2019_05001/itmconf_dictap2019_05001.html}, DOI={10.1051/itmconf/20192705001}, journal={ITM Web of Conferences}, author={Asante, Bismark Asiedu and Imamura, Hiroki}, year={2019}, month=jan, pages={05001} }

@misc{vaswani2023attentionneed,
      title={Attention Is All You Need}, 
      author={Ashish Vaswani and Noam Shazeer and Niki Parmar and Jakob Uszkoreit and Llion Jones and Aidan N. Gomez and Lukasz Kaiser and Illia Polosukhin},
      year={2023},
      eprint={1706.03762},
      archivePrefix={arXiv},
      primaryClass={cs.CL},
      url={https://arxiv.org/abs/1706.03762}, 
}

@misc{Alec,
      title={Robust Speech Recognition via Large-Scale Weak Supervision}, 
      author={Alec Radford and Jong Wook Kim and Tao Xu and Greg Brockman and Christine McLeavey and Ilya Sutskever},
      year={2022},
      eprint={2212.04356},
      archivePrefix={arXiv},
      primaryClass={eess.AS},
      url={https://arxiv.org/abs/2212.04356}, 
}

@misc{Junyi,
      title={SpeechT5: Unified-Modal Encoder-Decoder Pre-Training for Spoken Language Processing}, 
      author={Junyi Ao and Rui Wang and Long Zhou and Chengyi Wang and Shuo Ren and Yu Wu and Shujie Liu and Tom Ko and Qing Li and Yu Zhang and Zhihua Wei and Yao Qian and Jinyu Li and Furu Wei},
      year={2022},
      eprint={2110.07205},
      archivePrefix={arXiv},
      primaryClass={eess.AS},
      url={https://arxiv.org/abs/2110.07205}, 
}

@inproceedings{
rutunda2023kinyarwanda,
title={Kinyarwanda {TTS}: Using a multi-speaker dataset to build a Kinyarwanda {TTS} model},
author={Samuel Rutunda and Kleber Kabanda and Adriana Stan},
booktitle={4th Workshop on African Natural Language Processing},
year={2023},
url={https://openreview.net/forum?id=1gLgrqWnHF}
}

@inproceedings{
elamin2023multilingual,
title={Multilingual Automatic Speech Recognition for Kinyarwanda, Swahili, and Luganda: Advancing {ASR} in Select East African Languages},
author={Moayad Elamin and Yonas Chanie and Paul Ewuzie and Samuel Rutunda},
booktitle={4th Workshop on African Natural Language Processing},
year={2023},
url={https://openreview.net/forum?id=tuUHjowTKpC}
}

@incollection{rono2022development,
  title={Development of a Kiswahili Text-to-Speech System Based on Tacotron 2 and WaveNet Vocoder},
  author={Rono, Kelvin Kiptoo and Mwangi, Prof Elijah and others},
  booktitle={Development of a Kiswahili Text-to-Speech System Based on Tacotron 2 and WaveNet Vocoder: Rono, Kelvin Kiptoo| uMaina, Dr. Ciira wa| uMwangi, Prof Elijah},
  year={2022},
  publisher={[Sl]: SSRN}
}

@misc{kenya_wikipedia,
  author = {{wikipedia}},
  title = {Kenya - Climate},
  howpublished = {\url{https://en.wikipedia.org/wiki/Kenya\#Climate}},
  note = {Accessed: Sep. 04, 2020},
}

@inproceedings{commonvoice:2020,
  author = {Ardila, R. and Branson, M. and Davis, K. and Henretty, M. and Kohler, M. and Meyer, J. and Morais, R. and Saunders, L. and Tyers, F. M. and Weber, G.},
  title = {Common Voice: A Massively-Multilingual Speech Corpus},
  URL = {https://huggingface.co/datasets/mozilla-foundation/common_voice_11_0},
  note = {Accessed Jul. 05, 2024},
}

@misc{netlimiter,
  author = {{Netlimiter}},
  title = {NetLimiter},
  howpublished = {\url{http://www.netlimiter.com}},
  note = {Accessed Jul. 05, 2024},
}

@misc{kenya_isps,
  author = {{Broadband Speed Checker}},
  title = {ISPs in Kenya},
  howpublished = {\url{https://www.broadbandspeedchecker.co.uk/isp-directory/Kenya.html}},
  note = {Accessed: 2024-07-24}
}

@inproceedings{ahn2023performance,
  author = {Ahn, Hyunho and Chen, Tian and Alnaasan, Nawras and Shafi, Aamir and Abduljabbar, Mustafa and Subramoni, Hari and Panda, Dhabaleswar K.},
  title = {Performance Characterization of Using Quantization for DNN Inference on Edge Devices},
  booktitle = {Proceedings of the International Conference on Frontiers of Edge Computing (ICFEC)},
  year = {2023},
  month = {May},
  doi = {10.1109/icfec57925.2023.00009},
}

@article{Leroux2017TheCN,
  title={The cascading neural network: building the Internet of Smart Things},
  author={Sam Leroux and Steven Bohez and Elias De Coninck and Tim Verbelen and Bert Vankeirsbilck and Pieter Simoens and B. Dhoedt},
  journal={Knowledge and Information Systems},
  year={2017},
  volume={52},
  pages={791-814},
  url={https://api.semanticscholar.org/CorpusID:31871252}
}

@misc{statista2024,
  author = {{Statista}},
  title = {{Kenya mobile device vendor share 2018-2021}},
  howpublished = {\url{https://www.statista.com/statistics/1061470/market-share-held-by-mobile-phone-vendors-in-kenya/}},
  note = {Accessed: 2024-07-24}
}

@misc{gsmarena2024,
  author = {{Kaggle}},
  title = {{GSMArena Phone Dataset}},
  howpublished = {\url{https://www.kaggle.com/datasets/arwinneil/gsmarena-phone-dataset}},
  note = {Accessed: 2024-07-24}
}

@misc{dmusingu,
  author       = {},
  title        = {{dmusingu/WHISPER-SMALL-SWAHILI-ASR-CV-14 · Hugging Face}},
  howpublished = {\url{https://huggingface.co/dmusingu/WHISPER-SMALL-SWAHILI-ASR-CV-14}},
  month        = dec,
  year         = 2023,
  note         = {Accessed: 2024-08-16}
}

@misc{mbazaWhisper,
  author       = {},
  title        = {{mbazaNLP/Whisper-Small-Kinyarwanda · Hugging Face}},
  howpublished = {\url{https://huggingface.co/mbazaNLP/Whisper-Small-Kinyarwanda}},
  month        = jun,
  year         = 2023,
  note         = {Accessed: 2024-08-16}
}
